\documentclass[epj]{svjour}
\usepackage{graphics}
\begin{document}
\title{Forward dispersion relations and Roy equations in \mbox{\boldmath $\pi\pi$} scattering}
\author{R.~Kami\'nski\inst{1}, J. R. Pel\'aez\inst{2} \and F. J. Yndur\'ain\inst{3} 
}                     
%
%
\institute{Department of Theoretical Physics
Henryk Niewodnicza\'nski Institute of Nuclear Physics,
Polish Academy of Sciences,
31-242, 
Krak\'ow, Poland \and 
Departamento de F\'{\i}sica Te\'orica,~II
 (M\'etodos Matem\'aticos),
Facultad de Ciencias F\'{\i}sicas,
Universidad Complutense de Madrid,
E-28040, Madrid, Spain \and
Departamento de F\'{\i}sica Te\'orica, C-XI
 Universidad Aut\'onoma de Madrid,
 Canto Blanco,
E-28049, Madrid, Spain}
\date{Received: date / Revised version: date}
%
\abstract{
We review results of an analysis of $\pi\pi$ interactions in 
$S$, $P$ and $D$ waves for two-pion effective mass from threshold 
to about 1.4 GeV.
In particular we show a recent improvement of this analysis 
above the $K\bar K$ threshold using more data for phase shifts and 
including the $S0$ wave inelasticity from $\pi\pi \to K \bar K$. In addition,
we have improved the fit to the $f_2(1270)$ resonance and used a more
flexible $P$ wave parametrization above the $K\bar K$ threshold 
and included an estimation of the $D2$ wave inelasticity.
The better accuracy thus achieved also required a refinement of the Regge analysis
above 1.42 GeV. 
We have checked that the $\pi\pi$ scattering amplitudes obtained in this
approach satisfy remarkably well forward dispersion relations and 
Roy's equations.
} 
\maketitle
\section{Introduction}
\label{intro}
In a previous analysis \cite{PY05} a set of fits to different data sets on 
$\pi\pi$ scattering was presented together with a detailed description 
of the mathematical methods used in calculations.
Forward dispersion relations (FDR) were then used 
in order to test the correctness of the amplitudes thus constructed.
Remarkably, it was found that some of the 
very frequently used sets of phase shifts do not satisfy FDR 
below 1 GeV.
Thus FDR were shown to give strong constraints to fits which, when used later
 as a constraint, lead to an improved  and precise representation of
$\pi\pi$ scattering amplitudes below, roughly 1 GeV.
In the regions from 
about 1 GeV to 1.4 GeV there was still  
 some mismatch between the real part of the
amplitudes and the results of dispersive evaluations in \cite{PY05} (especially for
$\pi^0\pi^0$ scattering).

In a subsequent article \cite{KPY06}, in order to improve the agreement 
with the constraints given by FDR, we have reanalysed the
parametrizations of the $S0$ above $KK$ threshold,  the $D0$ wave 
and to a lesser extend the $P$ and $D2$ waves.
In the $S0$ wave we took into account systematically the elasticity data from 
the $\pi\pi \to K \bar K$  reaction
\cite{Hyams73,Grayer74,KLR,Hyams75,pipi.to.KK}, 
included in the fit more data on phase shifts above the $K\bar K$ threshold 
\cite{Hyams73,Grayer74,KLR,Hyams75} and used more a
flexible parametrization from 0.932 GeV to 1.4 GeV.
In the $D0$ wave we have used experimental data from \cite{Hyams73,Hyams75,Protop}, 
information on low energy parameters (the scattering length and slope)
and included in the fit the width and mass of the $f_2(1270)$ resonance
as given by the PDG \cite{PDG}. 
The result is that, for both $S0$ and $D0$ waves, we have obtained more 
accurate parametrizations 
with smaller errors compared to those in the previous approach \cite{PY05}.
In the $P$ wave we have exploited a more flexible 
parametrization between the $K\bar K$ threshold and 1.42 GeV and
in the $D2$ wave we have included its estimated inelasticity above 1 GeV.

This more accurate determination of the $\pi\pi$ amplitudes 
below 1.42 GeV allowed us to refine
the Regge analysis that had been used in \cite{PY05,Pelaez:2003ky}. 
This has been done by removing the degeneracy condition
$\alpha_{\rho}(0) = \alpha_{P'}(0)$ which thus modifies slightly the 
central values of the intercepts
$\alpha_{\rho}(0)$ (by $\sim 11\%$) and $\alpha_{P'}(0)$ (by $\sim 4\%$$), 
$
but yields smaller errors than those in \cite{PY05}.

We have found that the $\pi\pi$ amplitudes with the 
new parametrizations of phase shifts and 
inelasticities in the $S$, $P$ and $D$ waves
together with the just discussed small changes in $\alpha_{\rho}(0)$  and $\alpha_{P'}(0)$  
allow for much better fulfilment of FDR than in \cite{PY05}.
The biggest improve in $\chi^2$ (about $66\%$) 
is achieved for the forward $\pi^0\pi^0$ dispersion
relation and a smaller one (about $15\%$) for $\pi^0\pi^+$.
In the case of the forward dispersion relation for isopin 1 in the $t$-channel
a very tiny deterioration has been 
found ($\chi^2$ increased by about $26\%$), which is still acceptable,
since, considering all FDR together, there is
a considerable overall improvement in their fulfilment.
It is worth noting that 
this 
has been achieved despite the improved data fits
have smaller errors than in \cite{PY05}. 

We have also tested Roy equations, which, contrary to FDR, incorporate
s-t crossing, by
calculating the difference between the
real parts of the input amplitudes and those obtained from Roy's equations.
We have found that, on average, and up to almost the $KK$ threshold,
the deviation from zero is smaller that 1.05 times the
value of  the errors for the $S0$ wave, smaller that 1.2 for the $S2$ wave and 
smaller than 0.65 for the $P$ wave. 


\section{\mbox{\boldmath $S$}, \mbox{\boldmath $P$} and \mbox{\boldmath $D$} waves at higher energies but below 1.42 GeV}
\label{SPD}

In this section the main features of the
new paramaterizations of $S$, $P$ and $D$ waves  
between roughly the $K\bar K$ threshold and 1.42 GeV are presented. 
Details of each parametrization can be found in \cite{KPY06}.
Since the description of the $S2$ wave was not changed in \cite{KPY06}, 
any information on this wave is available in \cite{PY05}.


\subsection{The \mbox{\boldmath $S0$} wave}
\label{S0subs}

\begin{figure}
\resizebox{.7\textwidth}{16cm}{%
\hspace{-2.cm}\includegraphics{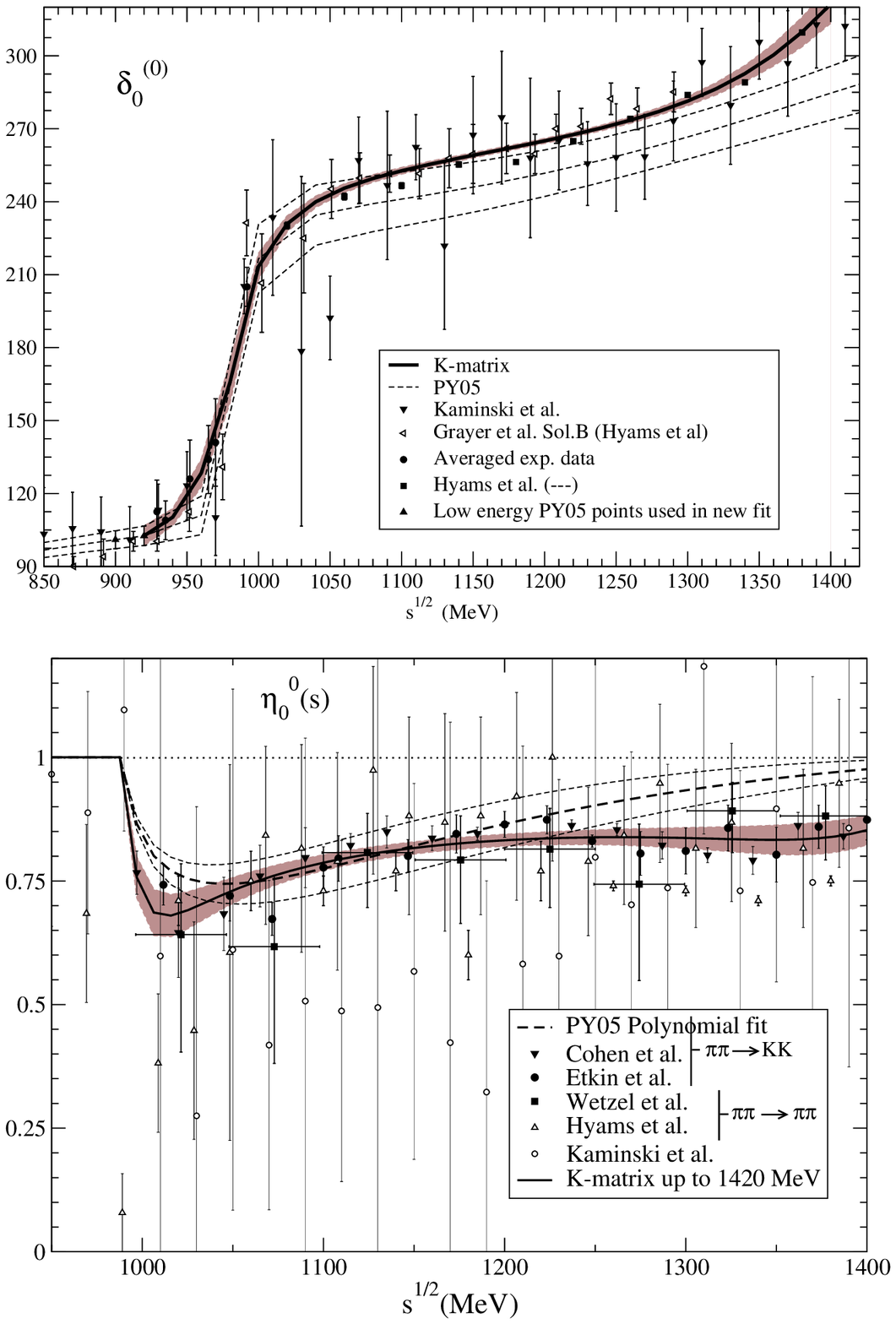}}

\vspace{-5cm}

\caption{Phase shifts and inelasticities of the $S0$ wave fitted using the
{\bf K}-matrix approach of \cite{KPY06} (solid lines). Dotted
lines for the results of \cite{PY05}.}
\label{S0KPY06}       
\end{figure}

In the present approach we obtain both the tangent of the phase shifts 
tan$\delta^0_0$ and the inelasticity 
$\eta^0_0$  above 0.932 GeV
as functions of {\bf K}-matrix elements
\begin{equation}
K_{ij}(s)=\frac{\mu\alpha_i \alpha_j}{M_1^2-s}+
\frac{\mu\beta_i\beta_j}{M_2^2-s}+\frac{1}{\mu}\gamma_{ij},
\end{equation}
 where $i,j = 1,2$ denote $\pi$ or $K$ respectively, and we set the mass scale
$\mu = 1$ GeV. All $\alpha_i, \beta_i$ and $\gamma_i$ are determined from the fit. Note that
 $M_1 = 0.9105 \pm 0.0070$ GeV simulates the left hand 
 cut of the {\bf K}-matrix located at $2\sqrt{M_K^2-m_{\pi}^2}= 0.952$ GeV and the
 pole at $M_2 = 1.324 \pm 0.006$ GeV is connected
 with $\delta^0_0$ passing through $270^o$.
 The parametrizations: of \cite{PY05} (below 0.932 GeV) and of \cite{KPY06}
 (above 0.932 GeV)  are matched at 0.932 MeV.
 In the fit all data on 
 phase shifts below and above the $K\bar K$ threshold \cite{Hyams73,Grayer74,KLR,Hyams75} 
 have been used simultaneously. 
 For $\eta^0_0$,  data from
 $\pi\pi \to K \bar K$ have been used together with data on $\pi\pi \to \pi\pi$
 \cite{Hyams73,Grayer74,KLR,Hyams75,pipi.to.KK}.
 The resulting fit yields $\chi^2/d.o.f = 0.6$ and can be seen in Fig. \ref{S0KPY06}. 

\subsection{The \mbox{\boldmath $D0$} wave}
\label{D0subs}

For this wave we proceeded by fitting simultaneously
 below and above the $K \bar K$ a parametrization
\begin{equation}
cot\delta_2^{(0)} = \frac{s^{1/2}}{2k^5}(M^2_{f_2}-s)m_{\pi}^2(B_0+B_1w(s))
\end{equation}
with $w(s) = \frac{\sqrt{s}-\sqrt{s_0-s}}{\sqrt{s}+\sqrt{s_0-s}}$, but 
using different $B_i$  and $s_0$ parameters above and below $KK$ threshold.
We also required both parametrizations to match at $\sqrt{s} = 2m_K$, 
thus eliminating one parameter.
In the present approach,
 the mass of the $f_2(1270)$ resonance $M_{f_2}$ was fixed to the PDG value \cite{PDG}.
The $B_i$ parameters have been obtained for those two energy regions from 
fits to experimental data points \cite{Hyams73,Hyams75,Protop} 
together with three other constraints: the width of the $f_2(1270)$ resonance from \cite{PDG},
plus the scattering length and the slope parameter calculated from the
Froissart-Gribov representation. 
The resulting $\chi^2/d.o.f = 0.65$.

The inelasticity is parametrized in the 
same way as in \cite{PY05} and fitted to the experimental data of ref.\cite{Hyams73,Hyams75,Protop}
\begin{equation}
\eta_2^{(0)}(s) = 1 - \epsilon \frac{k_2(s)}{k_2(M^2_{f_2})}
\end{equation}

\begin{figure}
\resizebox{.7\textwidth}{18cm}{%
\hspace*{-2cm}\includegraphics{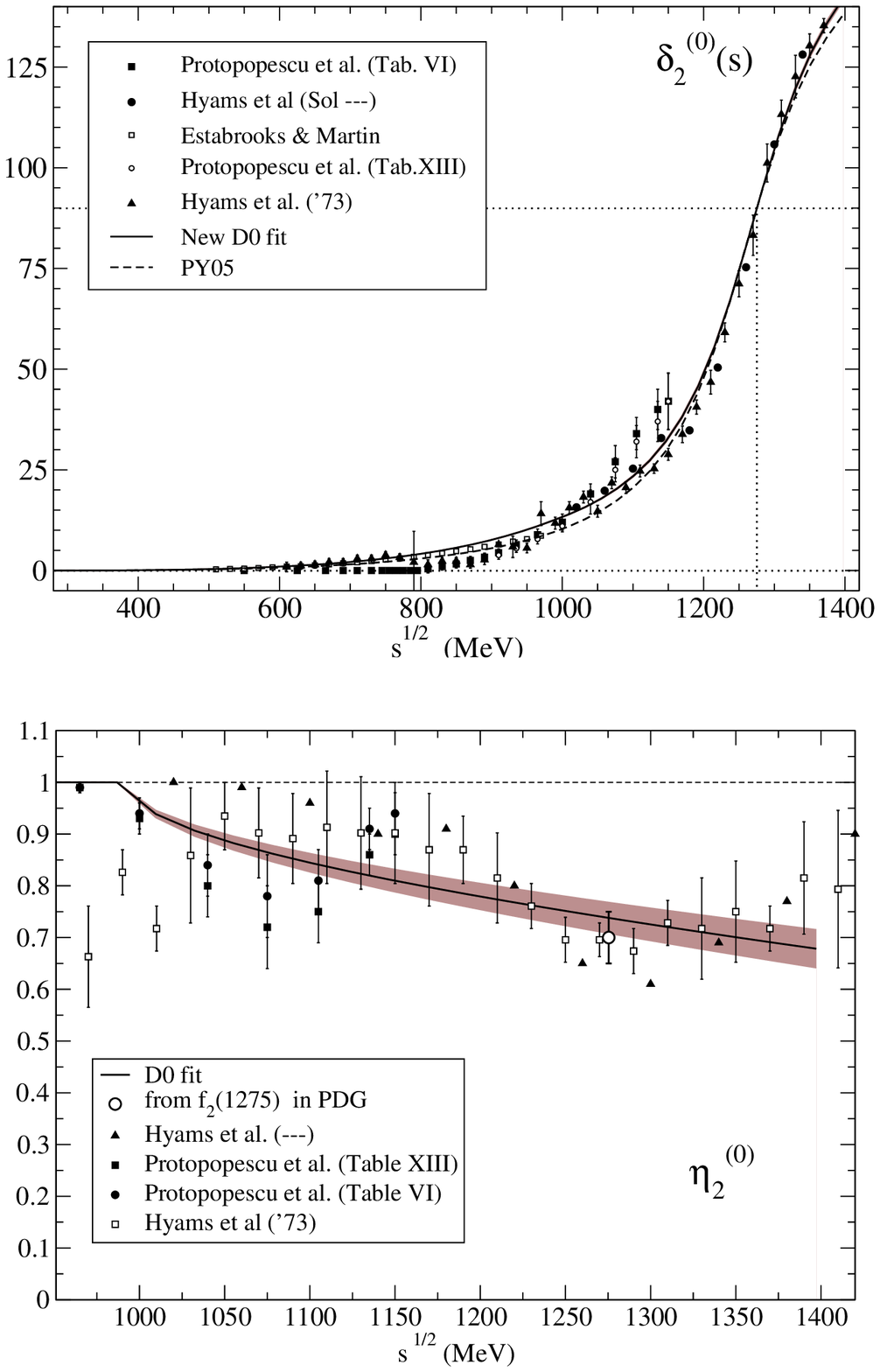}}

\vspace{-5.5cm}

\caption{The $D0$ wave phase shifts and inelasticities determined in \cite{KPY06} (solid lines) and in \cite{PY05} 
(dotted line - only for phase shifts). 
Dark areas denote the errors, which for the phase shifts 
have just the thickness
of the line.}
\label{D0}       
\end{figure}

Results of the fits for phase shifts and inelasticities 
are presented in Fig. \ref{D0}.


\subsection{The \mbox{\boldmath $P$} wave}
\label{Psubs}

\begin{figure}

\vspace{-.5cm}

\resizebox{0.5\textwidth}{15cm}{%
\hspace{-2cm}\includegraphics{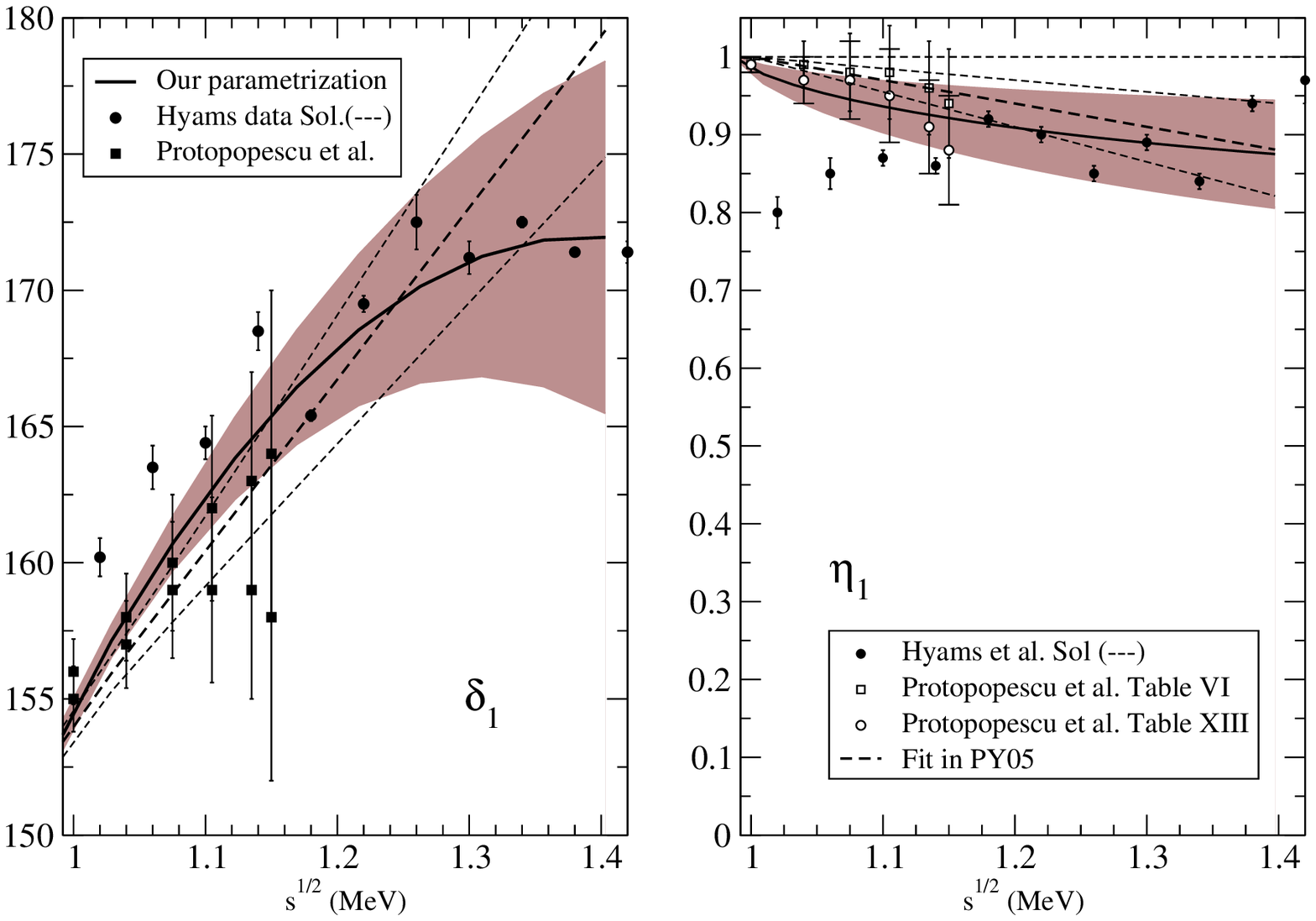}}

\vspace{-7.cm}

\caption{Fits to the $P$ wave phase shifts and inelasticity (solid lines).
Dark areas show the errors of our results.
The dotted lines represent results obtained in \cite{PY05}.}
\label{P}       
\end{figure}

In the $P$ wave, above the $K\bar K$ threshold 
we have used a more flexible parametrization than in \cite{PY05}:
\begin{equation}
\delta_1(s) = \lambda_0 + \sum\limits_{i=1}^2 \lambda_i(\sqrt{s/4m_K^2}-1)^i, 
\end{equation}
\begin{equation}
\eta_1(s) = 1 - \sum\limits_{i=1}^2 \epsilon_i(\sqrt{1-4m_K^2/s})^i,
\end{equation}
where $\lambda_0$ is fixed by the phase shift at $2m_K$ which is obtained from the 
fit to the pion form factor \cite{TY}.
We have then fitted data from \cite{Hyams75,Protop} 
obtaining $\chi^2/d.o.f.=0.6$ and 
$\chi^2/d.o.f.= 1.1$ for the phase shifts and inelasticity, respectively.
The results are presented in Fig. \ref{P}.

\subsection{The \mbox{\boldmath $D2$} wave}
\label{D2subs}

\begin{figure}

\vspace{-2cm}

\resizebox{.5\textwidth}{14cm}{%
\hspace{-2.cm}\includegraphics{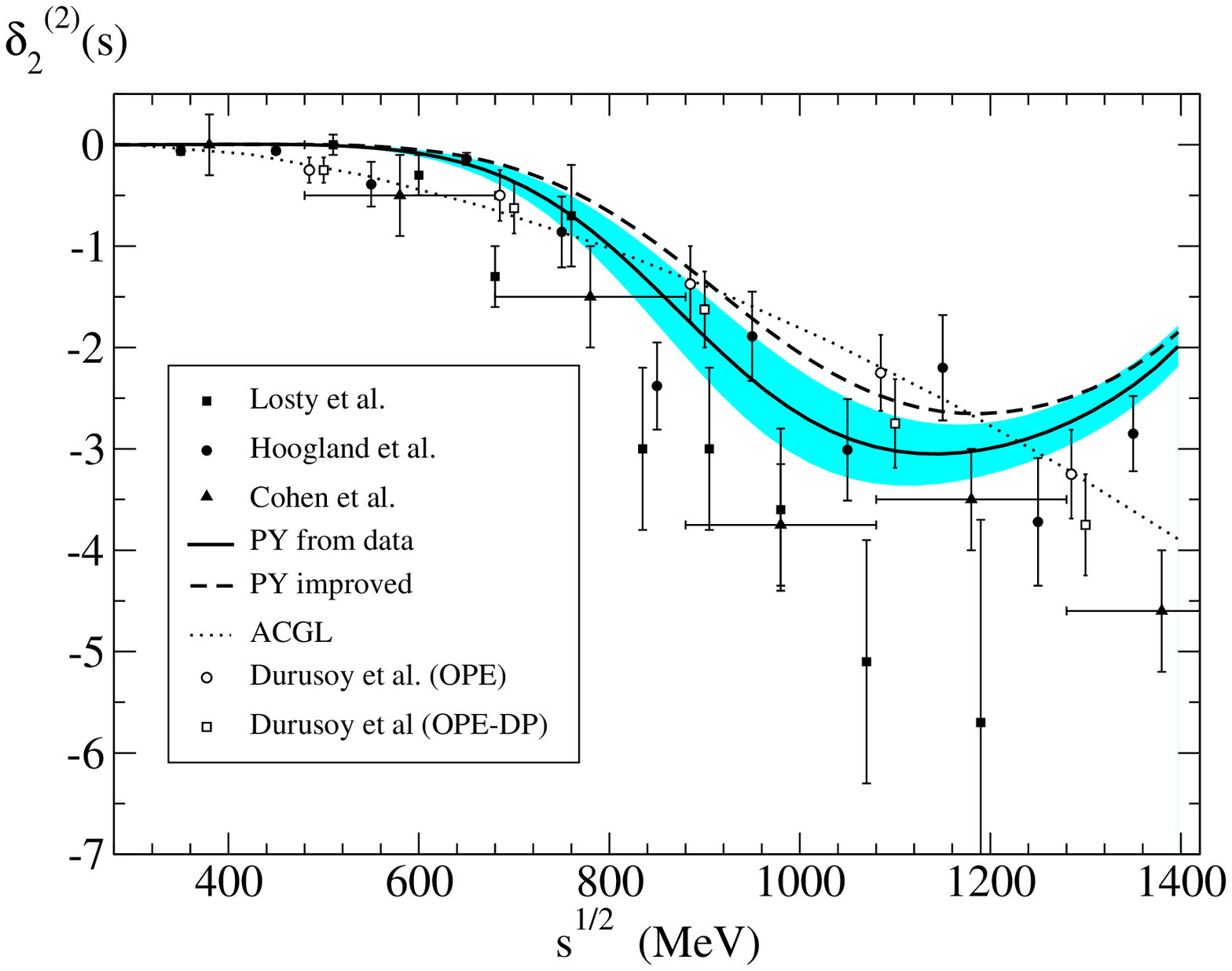}}

\vspace{-5.7cm}

\caption{Results of the fit to the 
$D2$ wave. The solid line denotes the fit to the experimental data only and 
the dashed one the fit to the data and FDR \cite{PY05}. 
For the data enclosed in the figure see references in \cite{KPY06}.}
\label{D2}       
\end{figure}

In the $D2$ wave we have used one 
single parametrization up to 1.42 GeV with four free parameters
\begin{equation}
cot\delta_2^{(2)} = \frac{s^{1/2}}{2k^5}\left(B_0+B_1w(s)+B_1w(s)^2\right)\frac{m_{\pi}^4s}{4(m_{\pi}^2+\Delta^2)-s},
\end{equation}
where $\Delta$ fixes zero of the phase shift near the $\pi\pi$ threshold.
Since the data on this wave are not accurate 
we have added one more constrain using the scattering length calculated from
the Froissart-Gribov representation \cite{PY05}.
As a result we have obtained the fit presented in Fig. \ref{D2}.

The lack of experimental data  on inelasticity led us to estimate it from a model 
(see ref. \cite{KPY06}) writing  
\begin{equation}
\eta^{(2)}_2 = 1-\epsilon(1-\hat s/s)^3, 
\end{equation}
with $\sqrt{\hat s} = 1.05$ GeV and $\epsilon = 0.2 \pm 0.2$. 
The inelasticity is very small and even negligible below 1.25 GeV.


\section{Regge parametrization}
\label{Reggeparam}

\begin{figure}
\resizebox{.7\textwidth}{15cm}{%
\hspace{-2cm}\includegraphics{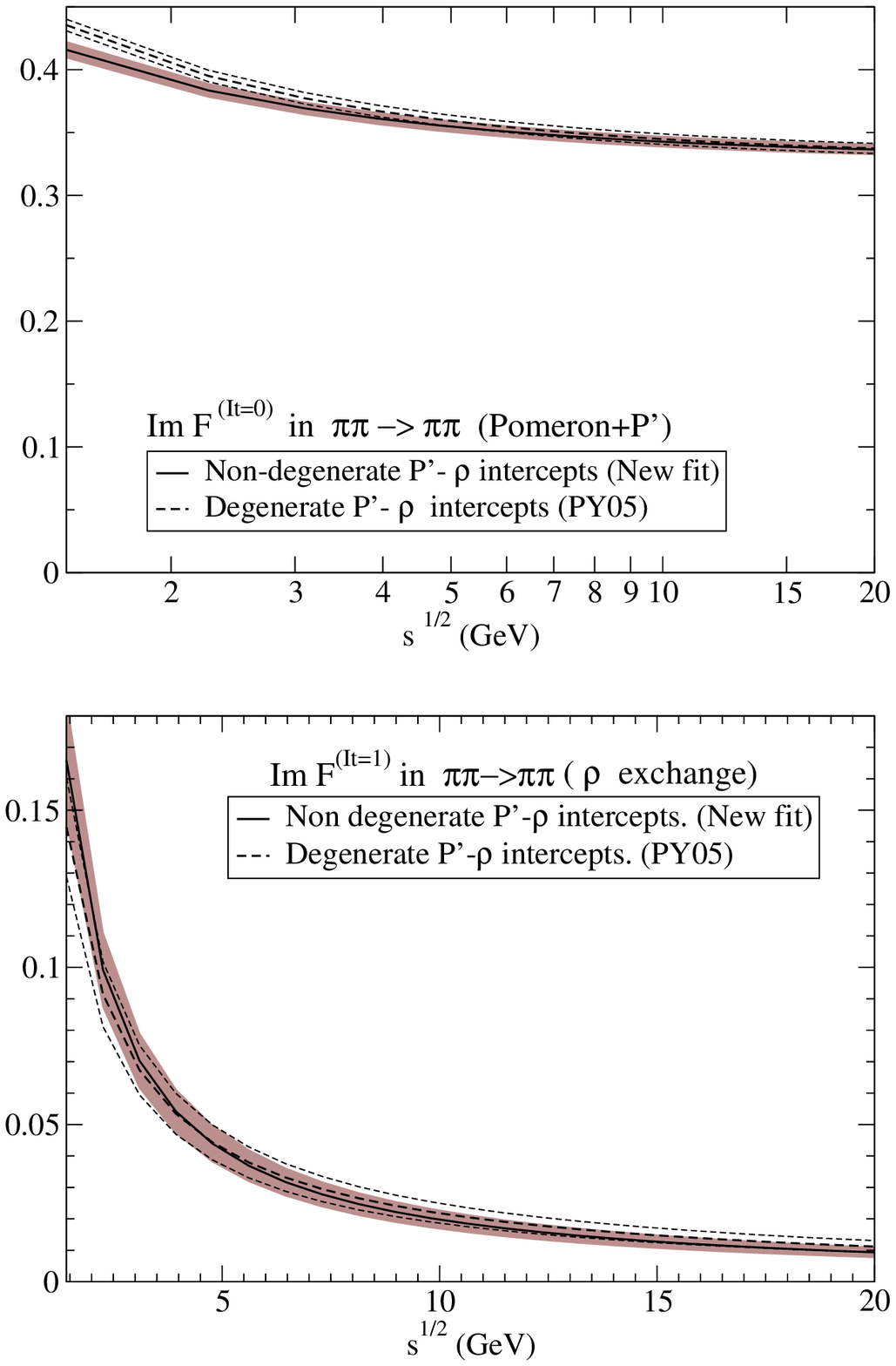}}

\vspace{-4.5cm}

\caption{The scattering amplitudes calculated 
with "exact degeneracy"  (broken lines) and 
without this condition (solid lines). Dark bands stand for uncertainties.} 
\label{Regge}       
\end{figure}

In the analysis of \cite{PY05,Pelaez:2003ky} 
the fits were made with the
assumption of "exact degeneracy" of the intercept parameters
$\alpha_{\rho} = \alpha_{P'}$  for $\rho$ and $f_2$ exchange.
In our new approach this degeneracy has been lifted. As a consequence, 
there was a very small change in  
the high energy behaviour of scattering amplitudes (especially a little for higher energies) 
but, as can be seen in next section,  
even such a small change could be significant given
the level of precision achieved in our FDR calculation. 
The energy dependence of the new scattering 
amplitudes after eliminating the degeneracy is seen in Fig. \ref{Regge}.

\section{Implementation of forward dispersion relations}
\label{UseFDR}

\begin{figure}

\vspace{-1cm}

\resizebox{.9\textwidth}{19cm}{%
\hspace{-2.5cm}\includegraphics{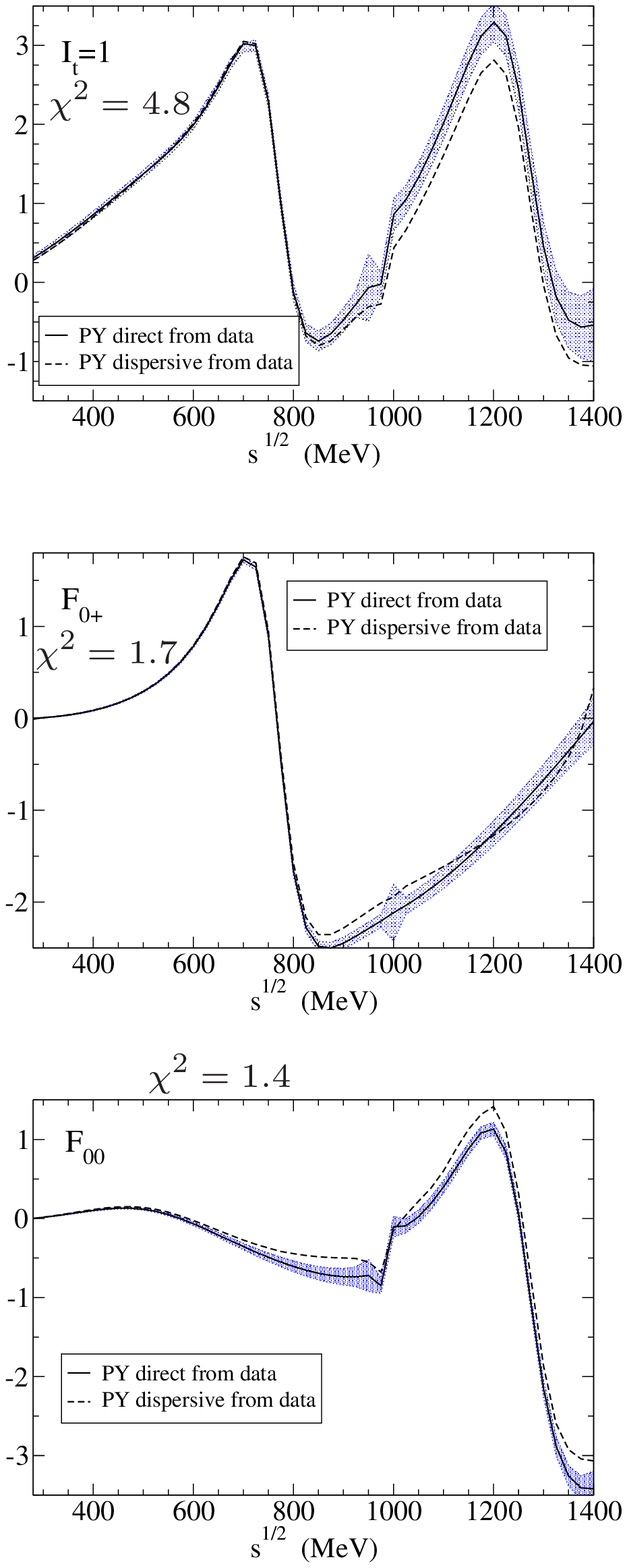}}

\vspace{-2.3cm}

\caption{The $\pi^0\pi^0$, $\pi^0\pi^+$ and t-channel 1 forward
dispersion relations 
described in previous analysis \cite{PY05}. 
The values of the $\chi^2$ denote averaged values over all 25 points chosen in the 
energy range from the $\pi\pi$ threshold to 1.42 GeV.}
\label{FDRPY05}       
\end{figure}

\begin{figure}

\vspace{-1cm}

\resizebox{.9\textwidth}{19cm}{%
\hspace{-2.5cm}\includegraphics{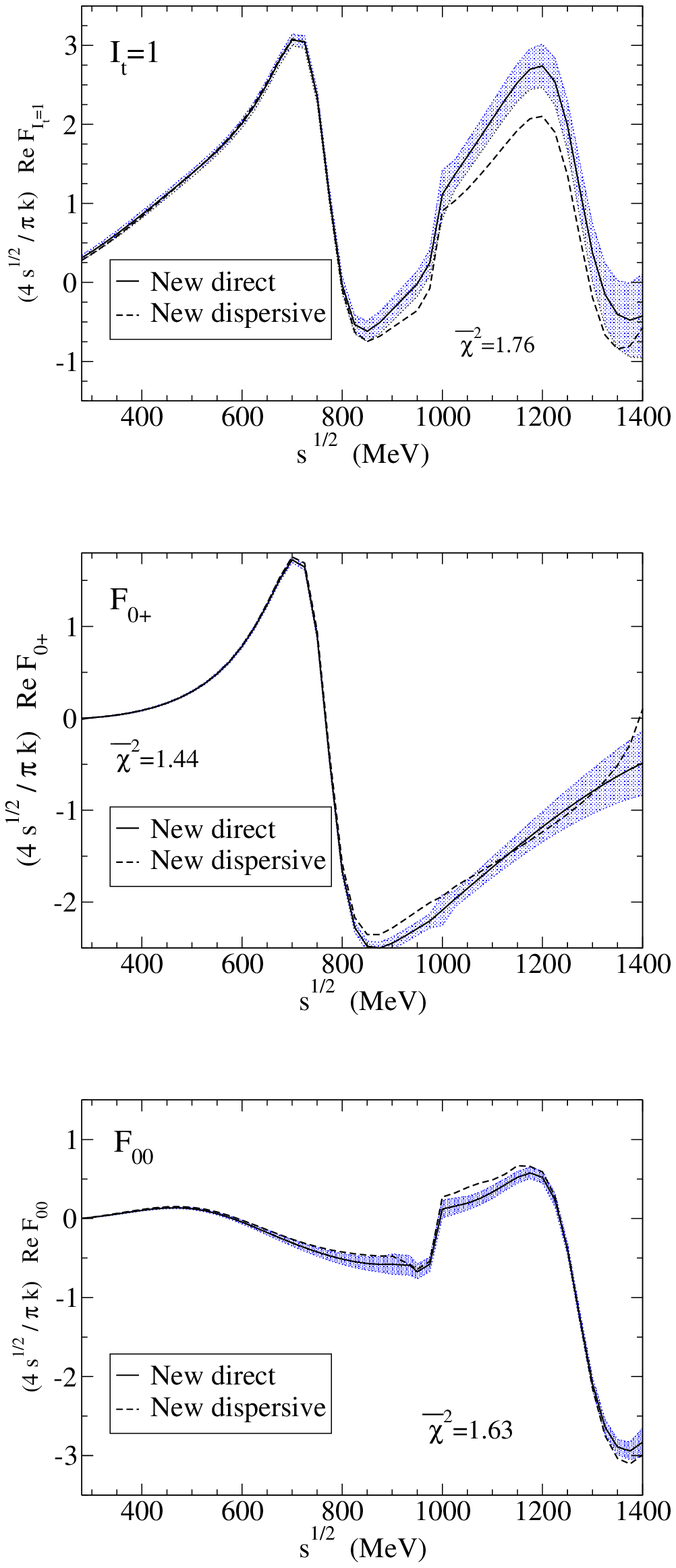}}

\vspace{-2.3cm}

\caption{Dispersion relations for new $S$, $P$ and $D$ waves described 
in this paper. The $\chi^2$ definition as in Fig.
\ref{FDRPY05}}
\label{FDRKPY06}       
\end{figure}

The $S$, $P$ and $D$ waves presented in Section \ref{SPD} 
together with the improved Regge description in the previous section 
have been examined in the same way as in \cite{PY05}, by checking the FDR's,
but without 
imposing  them as constraints. 
Thus, in Fig. \ref{FDRPY05} we present the
results from the amplitudes in \cite{PY05} obtained from 
fits to data. 
In contrast, in Fig. \ref{FDRKPY06} 
we show results using the improved  
fits given in \cite{KPY06} that we are reviewing
here.
The $F_{00}$, $F_{0+}$ and $I_t=1$ names used 
in Figs \ref{FDRPY05} and \ref{FDRKPY06} correspond to the FDR's for the 
$\pi^0\pi^0$, $\pi^0\pi^+$ and $t$-channel isospin 1 scattering amplitudes,
whose full mathematical expressions can be found in \cite{PY05} and 
\cite{KPY06}. The word "dispersive" denotes results obtained from the integrals in the
 FDR's  whereas
"direct" means the real parts evaluated directly from
parametrizations.

\begin{table}
\caption{Comparison of averaged $\chi^2$ for different FDR 
obtained in previous analysis \cite{PY05} and in presented one (new $\delta, \eta$ and new Regge) in two energy ranges. 
Numbers correspond to fits to experimental data only (without constraints from FDR).}
\label{tab:1}       
\begin{tabular}{cccc}
results of \cite{PY05} & new $\delta, \eta$ &  new Regge & Energy range\\
\hline\noalign{\smallskip}
\multicolumn{3}{c}{for $\pi^0\pi^0$ dispersion relations:} &  \\
\hline\noalign{\smallskip}
3.8 & 1.52 & 1.41  &  $s^{1/2}< 930$ MeV \\
4.8 & 1.76 & 1.63  & $s^{1/2}< 1420$ MeV \\
\hline\noalign{\smallskip}
\multicolumn{3}{c}{for $\pi^0\pi^+$ dispersion relations:} & \\
1.7 & 1.75 & 1.60  & $s^{1/2}< 930$ MeV\\
1.7 & 1.60 & 1.44  & $s^{1/2}< 1420$ MeV\\
\hline\noalign{\smallskip}
\multicolumn{3}{c}{for $I_t = 1$ scattering amplitudes:} & \\
0.2 & 0.57 & 0.32  & $s^{1/2}< 930$ MeV\\
1.4 & 2.32 & 1.76  & $s^{1/2}< 1420$ MeV\\
\end{tabular}
\end{table}

We provide in Table \ref{tab:1}  the FDR's averaged $\chi^2$ obtained
over the range from threshold up to 930 MeV or 1420 MeV.
Note that
the modifications in the $S$ and  $P$ waves above $KK$ threshold, 
as well as of the $D$ wave, lead to significant 
improvement of accuracy in the FDR 
for the $\pi^0\pi^0$ scattering amplitudes 
when compared with the previous results in \cite{PY05}. 
The final decrease of the 
$\chi^2$ for $\pi^0\pi^+$ is also due to the 
influence of the new Regge amplitude. 
Note that in the $I_t=1$ case there is
a tiny deterioration despite a significant 
 $\chi^2$ decrease due to the Regge part.


\section{Tests using Roy's equations}
\label{TestRoy}
We here present an advance of our ongoing analisys where we test
our new $\pi\pi$ scattering
 amplitudes using Roy's equations \cite{Roy:1971tc,Ananthanarayan:2000ht,kll2003}:
\begin{equation}
    \begin{array}{ll}
\!    \mbox{Re } f_{\ell}^{I}(s) =
        a_{0}^{0}\delta_{I0} \delta_{\ell 0} + a_{0}^{2}\delta_{I2}
        \delta_{\ell 0} \,\,+ \\
     \displaystyle
\!\!(2a_{0}^{0}-5a_{0}^{2})     
    (\delta_{I0}\delta_{\ell 0}\!+\!
        \frac{1}{6}\delta_{I1}\delta_{\ell 1} 
        \!\!-\!\! \frac{1}{2}\ \delta_{I2}\delta_{\ell 0})\frac{s\!-4{\mu}^2}{12{\mu}^2}+\\
     
\!      \!\displaystyle \sum\limits_{I'=0}^{2}
        \displaystyle \sum\limits_{\ell'=0}^{1}
     \;\;\,\,\!\!-
        \!\!\!\!\!\!\!\!\!\displaystyle \int \limits_{4\mu^2}^{s_{max}}\!\!\!\! ds'
     K_{\ell \ell^\prime}^{I I^\prime}(s,s') \mbox{Im}f_{\ell'}^{I^\prime}\!(s')+ 
     d_{\ell}^{I}(s,s_{max}),
\end{array}
 \label{royeq}
\end{equation}
where 
\begin{equation}
f_{\ell}^{I}(s)=\sqrt{\frac{s}{s-4\mu^2}}\frac{1}{2i}
    \left(\eta_{\ell}^{I}e^{2i\delta_{\ell}^{I}}-1\right),
\end{equation}
with $a_{0}^{0}$ and $a_{0}^{2}$ being the $S0$ and $S2$ scattering lengths, 
$K_{\ell \ell^\prime}^{I I^\prime}(s,s')$ known kernels and
$d_{\ell}^{I}(s,s_{max})$ the so called driving terms.
In our calculations we have chosen $s_{max} = 103 m_{\pi}^2$.

\begin{figure}
\resizebox{.85\textwidth}{16.5cm}{%
\hspace{-2cm}\includegraphics{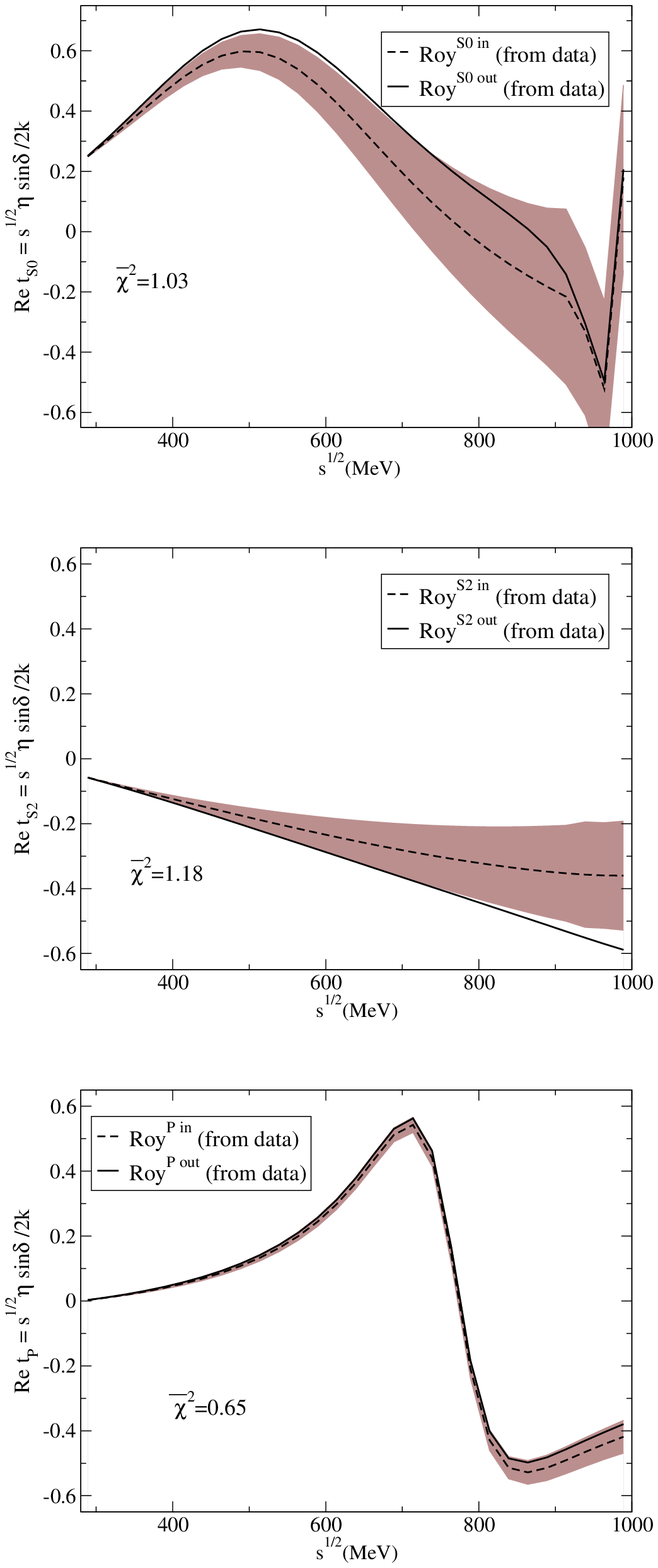}}

\vspace{-1cm}

\caption{Differences between real parts 
of amplitudes calculated directly from amplitudes and those from the integral
representation of Roy's equations. 
The notation is explained in the Section \ref{TestRoy}.}
\label{Roy}       
\end{figure}

In Fig. \ref{Roy} we show the real
part of the $S0$, $S2$ and $P$ partial waves obtained from 
Eq.(\ref{royeq}), (continuous line, called Roy$^{out}$)
versus the real part obtained directly from 
our parametrizations, 
(dashed line, called called Roy$^{in}$).

The agreement is remarkable, taking into account the uncertainties (the
dark areas in Fig. \ref{Roy}).
Furthermore, the agreement is even more impressive,
taking into account that we have not imposed
any constraints from FDR or Roy's equations themselves
and that the amplitudes come just from fits to data 
(that is why they are labeled ``from data'' in the Figure). Moreover,
 we use the new
$S$, $P$ and $D$ waves described in Section \ref{SPD} and the Regge model with 
different intercepts $\alpha_{\rho}(0)$ and $\alpha_{P'}(0)$, and
all of them have experimental errors even smaller 
than those of \cite{PY05}.  

\section{Conclusions}
\label{Conclusions}

The results reviewed here indicate 
that the improvement in the fits to data 
in the $S$ and $P$ waves above $KK$ threshold and the $D$ wave
described in Section \ref{SPD} together with a slight improvement in Regge
trajectories, allowing for non $\rho-f$ degeneracy,
also improves the fulfillment of forward dispersion relations.
Despite the smaller errors of those amplitudes,
 the averaged $\chi^2$ is indeed lower than in previous analysis \cite{PY05}.
We have shown that those amplitudes fulfill also 
quite well Roy's equations, and therefore, crossing symmetry, up to roughly 1 GeV.

\begin{figure}
\resizebox{.85\textwidth}{16.5cm}{%
\hspace{-2cm}\includegraphics{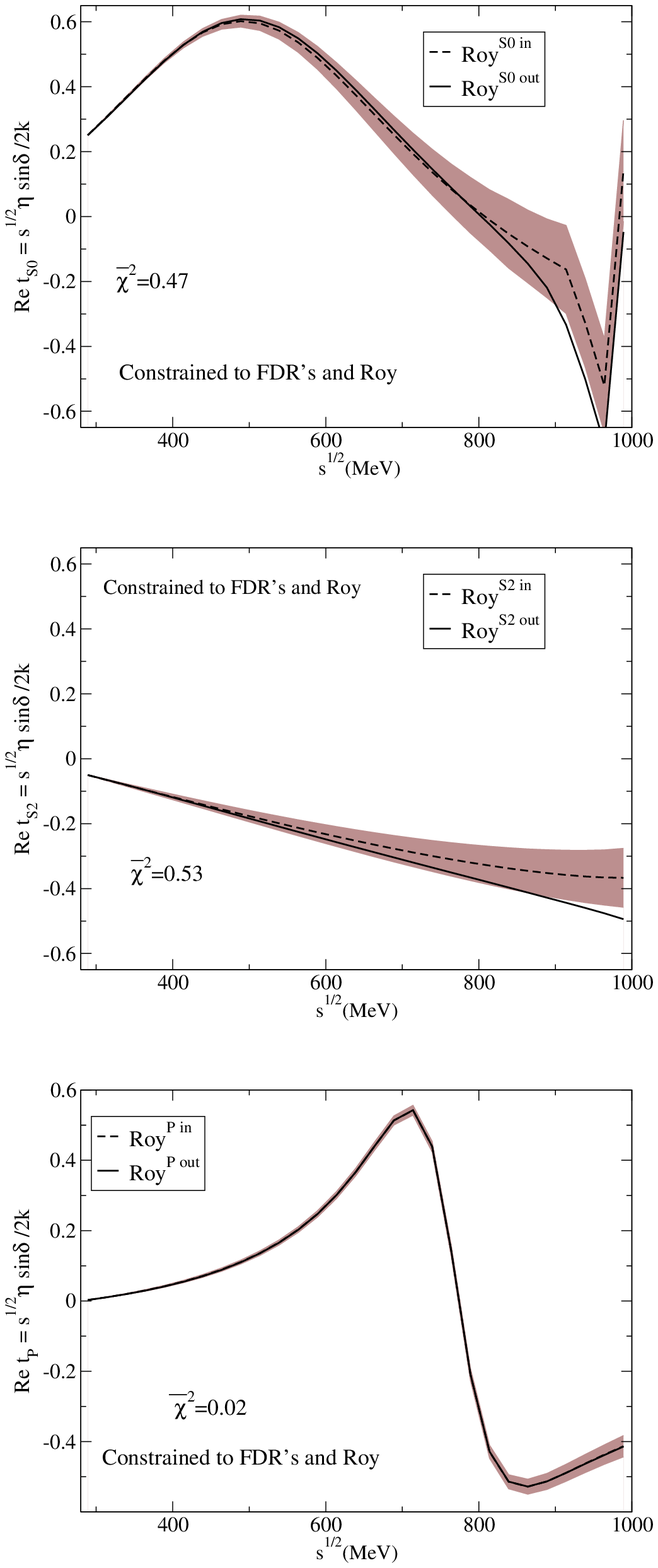}}

\vspace{-1cm}

\caption{As in Fig. \ref{Roy} but 
for data fits constrained by FDR and Roy's equations.}
\label{Royimp}       
\end{figure}

Following, however, 
the analysis done in \cite{PY05} 
one can think about a wider implementation of FDR including them into 
the fits together with the already fitted experimental data. 
We report briefly on our progress in this approach,
where, for the moment, we allow for
a variation of all the amplitude parameters except,  
the $P$ wave above $KK$ threshold and just the $\alpha$ and $\beta$  parameters
in the ${\bf K}-$matrix.
Although our results are just preliminary,
we already noticed significant decreases of the averaged $\chi^2$ for all three FDR. 
The preliminary values for $F_{00}$ decreased
 from 1.63 to 0.42, for $F_{0+}$ 
changed slightly from 1.44 to 1.48 and for $I_t=1$ decreased
from 1.76 to 0.89.
The more spectacular improvement, however, 
has been noticed in the Roy equations.
Preliminary averaged $\chi^2$ values 
decreased from 1.03 to 0.47 for the $S0$ wave, 
from 1.18 to 0.53 for the $S2$ wave and from 0.65 to 0.02 for
the $P$ wave. 
This improvement can be clearly seen when 
comparing Figs \ref{Roy} and \ref{Royimp}.
Imposing the constrains from Roy equations and particularly 
from FDR, which is much stringent,
 leads to modifications
in the $S$, $P$ and $D$ waves by less than $1\sigma$ 
(with the exception for $D2$ wave where the empirical fit changed by 
$\sim 1.3\sigma$) and to negligible modifications in all other waves.
The resulting uncertainties are also significantly reduced 
with this approach as can be seen, just for Roy equations, 
in Figs \ref{Roy} and \ref{Royimp}.
At present we are finishing the determination of the final parameters
and their uncertainties.

\end{document}